\documentclass[fleqn, twocolumn]{wlscirep} 
\usepackage[utf8]{inputenc} 
\usepackage{graphicx}
\usepackage{hyperref}
\usepackage{epstopdf}
\usepackage{float}
\usepackage{amsmath}
\usepackage{amsfonts}
\usepackage{bm}
\usepackage{titlesec}

\graphicspath{{Figures/}} 

\title{Searching High Temperature Superconductors with the assistance of Graph Neural Networks}

\normalsize

\author[1,2]{Liang Gu}  
\author[1,2]{Yang Liu}
\author[3,*]{Pin Chen}
\author[1,2,*]{Haiyou Huang}
\author[4]{Ning Chen}
\author[5]{Yang Li}
\author[3]{Yutong Lu}
\author[1,2,*]{Yanjing Su}

\affil[1]{
    Beijing Advanced Innovation Center for Materials Genome Engineering, University of Science and Technology Beijing, Beijing, 100083, China
}
\affil[2]{
    Institute for Advanced Materials and Technology, University of Science and Technology Beijing, Beijing, 100083, China
}
\affil[3]{
    National Supercomputer Center in Guangzhou, School of Computer Science and Engineering, Sun Yat-sen University, Guangzhou, 510006, China
}
\affil[4]{
    School of Materials Science and Engineering, University of Science and Technology Beijing, Beijing, 100083, China
}
\affil[5]{
    Department of Engineering Science and Materials, University of Puerto Rico, Mayaguez, Puerto Rico, 00681-9000, USA
}
\affil[*]{
    E-mail:
    huanghy@mater.ustb.edu.cn,
    chenp85@mail.sysu.edu.cn,
    yjsu@ustb.edu.cn
}

\begin{abstract}
Predicting high temperature superconductors has long been a great challenge. 
A major difficulty is how to predict the transition temperature ($T_{\rm{c}}$) of superconductors. 
Recently, progress in material informatics has led to a number of machine learning models predicting $T_{\rm{c}}$, which greatly improves the efficiency of prediction.
Unfortunately, prevailing models have not shown adequate physical rationality and generalization ability to find new high temperature superconductors, yet. 
In this work, in order to give a trustable prediction on the unexplored materials, 
we built a bond-sensitive graph neural network (BSGNN), which is optimized to process the information of chemical bond and electron interaction in the crystal lattice,
to predict the $T_{\rm{c}}$ maximum ($T_{\rm{c}}^{\rm{max}}$) of each type of superconducting materials. 
On the basis of the domain knowledge considered in the data preparation and algorithm design,
our model revealed a relevance between the $T_{\rm{c}}^{\rm{max}}$ and chemical bonds. 
The results indicate that shorter bond length is favored by high $T_{\rm{c}}$, which is in accordance with previous human experience. 
Moreover, it also shows that some specific chemical elements are favored by high $T_{\rm{c}}$, which is beyond what human experts already knew. 
It gives a convenient guidance for searching high temperature superconductors in materials database, by ruling out the materials that could never have high $T_{\rm{c}}$. 
\end{abstract}


\begin{document}

\flushbottom
\maketitle

\thispagestyle{empty}

\section*{1. Introduction}

    As the practical application of superconductors is severely constrained by the low working temperature, hunting for new superconductors with higher transition temperature ($T_{\rm{c}}$) is a long-cherished dream for generations. 
    However, finding new high temperature superconductor (HTS) is still a significant challenge.\cite{insuspense, Towardscompletetheory2006, Chu2015, Geballe2015, Norman2016, Stewart2017, Hirsch2015}     
    The difficulty lies in how to get an accurate and efficient prediction of $T_{\rm{c}}$, especially for the unexplored materials. 
    As many HTS are unconventional superconductors, the superconducting mechanism of which is still unclear, it is unfeasible to predict their $T_{\rm{c}}$ theoretically. 
    Meanwhile, empirical laws of the variation of $T_{\rm{c}}$ were supposed to give the guidance for predicting $T_{\rm{c}}$.
    Unfortunately, it has been found that the $T_{\rm{c}}$ can be affected by a good number of factors.\cite{Uemura1989, Uemura1991, Nakamura1996, planebuckling, Homes2004, Wilson2006, specificheat, anionheight, Stewart2011, Hashimoto2012, Lee2012, Pines2013, Bozovic2016, Wu2016, apicaloxygen, Cao2018}
    Several key factors are highly interrelated (varies with each other non-linearly), resulting in intricate dependencies between the $T_{\rm{c}}$ and those factors. 
    Not to mention that the data of those factors is often hard to obtain experimentally. 
    Therefore, it calls for a feasible method to reveal the patterns in the variation of $T_{\rm{c}}$, making use of abundant data readily available. 
    For the materials exploration of HTS, the input data should preferably be the chemical composition and crystalline structure, instead of any experimental data of physical parameters. 

    In this instance, machine learning (ML) has recently become a principal tool for searching HTS. 
    In the latest few years, plenty of ML models were developed to predict the $T_{\rm{c}}$ of superconductors as well as the existence of superconductivity.\cite{Stanev2018, Matsumoto2019, Functional, ATCNN, hydrides, Liu2020, Konno2021, BCSinspired, Eliashberg} 
    For example, models predicting $T_{\rm{c}}$ were trained with 10000+ data mainly derived from the database SuperCon\cite{Supercon}.
    Then the ML models were employed to identify HTS from the materials in open access databases such as Inorganic Crystallographic Structure Database (ICSD) \cite{ICSD} and Crystallography Open Database (COD)\cite{COD}. 
    So far, several algorithms, including random forest (RF)\cite{Stanev2018}, atom table convolutional neural network (ATCNN) \cite{ATCNN, Konno2021} and convolutional gradient boosting decision tree (ConvGBDT)\cite{CGBDT}, have achieved good predictive scores on the test data. 

    However, a model yielding good predictive score alone is not enough for the mission of searching new HTS. 
    One could get a lot of models giving accurate prediction on the test data (explored superconductors), 
    but many of them do not necessarily have adequate generalization ability and physical rationality, 
    thereby being bad in predicting unseen data (unexplored materials). 
    Although earlier state-of-the-art models have achieved higher $R^{2}$,\cite{ATCNN, Stanev2018}
    they didn't considered the crystalline structure information, which is vital for the superconductivity. 
    As a result, high scoring models so far have not successfully found a new HTS (experimentally verified), yet.     
    Moreover, different models gave inconsistent results when proposing candidates of HTS, implying that most models, perhaps all of them, are not as reliable as it seems. 
    Therefore, in the next stage, the crux is how to take into account essential domain knowledge during data preparation and algorithm design, so as to get a more trustable prediction on HTS.     
    For the mission of searching new HTS, a trustable ML model should have universal generalization ability (being accurate in predicting all kinds of materials) and explicit interpretability (indicating the key factors affecting $T_{\rm{c}}$ with physical meaning). 
    Otherwise, the model is nothing but just only a mathematical game. 
    
    In this study, we propose a way to give reliable prediction of $T_{\rm{c}}$ maximum ($T_{\rm{c}}^{\rm{max}}$) of each type of superconducting materials. 
    An graph neural network (GNN) model is utilized not only to predict $T_{\rm{c}}^{\rm{max}}$, but also to unveil the key factors affecting $T_{\rm{c}}$. 
    Being good at processing structure information and interaction information, GNN models for general purpose in materials science have shown impressive performance in predicting various material properties.\cite{CGCNN, Chen2019, ALIGNN, Isayev2015}     
    It is notable that GNN model has already achieved good performance in predicting conventional superconductivity,\cite{BCSinspired} 
    but still not in predicting HTS. 
    Despite of the abundance of architectures of GNN algorithm\cite{architectures}, it is unclear yet that which architectures are suitable and competent for the problem of unconventional superconductivity. 
    Here, we focus on the information of chemical bond and electron interaction in the crystal lattice, which is known as crucial factors influencing the superconductivity. 
    We develop a bond-sensitive graph neural network (BSGNN) algorithm, so as to adapt for the problem of superconductivity. 
    Three modules, called nearest-neighbors-only graph representation (NGR), communicative message passing (CMP) and graph attention (GAT), respectively, is integrated into our BSGNN algorithm.     
    The BSGNN model is used to search potential HTS materials in ICSD. 
    The prediction given by the model is further screened manually, resulting in a list of promising HTS candidates.
    We also discuss the relevant between the $T_{\rm{c}}$ and the chemical bonds. 
    It suggests that the bond length and the chemical elements composing the bond can be seen as the gene of HTS.

\section*{2. Methodology} 

    \subsection*{2.1 Data preparation} 
        
        Our GNN models were trained with the data of 612 superconductors. 
        The crystalline structures of those superconductors are derived from ICSD\cite{ICSD} (up to 2019), and the $T_{\rm{c}}$ values of them are mainly derived from SuperCon\cite{Supercon} (up to 2014). 
        Some compressed hydrides and some superconductors discovered recently were added into the dataset-SuperCon.
        The intersection set of dataset-SuperCon and dataset-ICSD has 1526 materials. 
        In those materials, we took the data that $T_{\rm{c}}$ $>$ 5 K (612 entries) as the input data for our GNN model. 
        
        Please note those data of $T_{\rm{c}}$ are mostly the highest $T_{\rm{c}}^{\rm{max}}$ value in their own material families. 
        Therefore, the target variable is actually the $T_{\rm{c}}^{\rm{max}}$ of each type of materials, instead of the $T_{\rm{c}}$ of each chemical composition. 

        The 612 input data were randomly (but not fully random) divided into train/test sets in a proportion of 9:1. 
        Before the dataset splitting, the materials family of each data was labeled. 
        During the dataset splitting, the ratio of each materials family was maintained for both the train set and test set. 
        As a result, the data distribution is better than the case of complete random data splitting. 
        
    \subsection*{2.2 Algorithm design} 

        
        \begin{figure}[!tb]
            \centering
            \includegraphics[width = 6.4 cm]{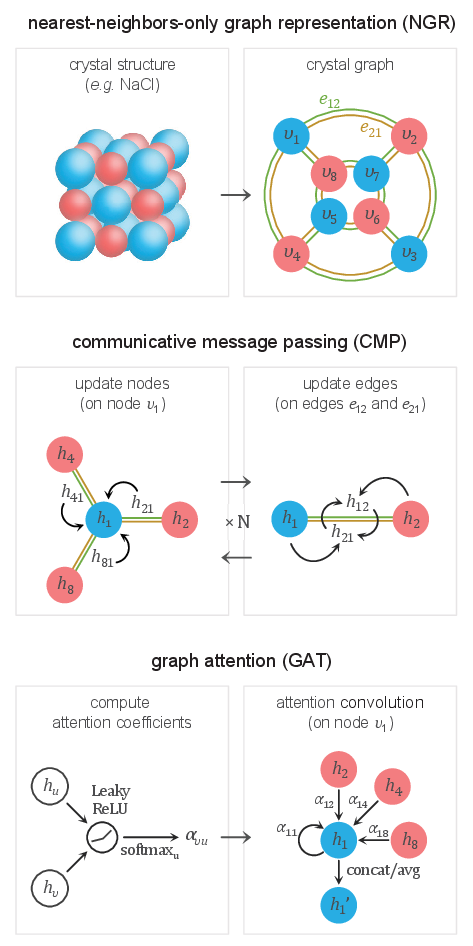}  
            \caption{
            Three key modules in BSGNN: nearest-neighbors-only graph representation (NGR) module, communicative message passing (CMP) module, and graph attention (GAT) module. $\upsilon_{n}$ and $e_{uv}$ represent the nodes and edges, respectively. $h_{n}$ and $h_{uv}$ represent the node features and edge feature, respectively. $\alpha_{uv}$ is the attention coefficients. $n$, $u$, $v$ are indices of nodes. 
            }
            \label{f1}
        \end{figure}
        
        
        It is well known that the crystalline structure is crucial for the superconductivity. 
        How to deal with the information of interaction between atoms or electrons in the crystal lattice, or to say, the bond information, is vital for a ML model predicting superconductors. 
        So, three modules, a nearest-neighbors-only graph representation (NGR) module, a communicative message passing (CMP) module\cite{MATGEN}, and a graph attention (GAT) module\cite{GAT}, were employed to capture the bond information. 
        
        As shown in Fig.~\ref{f1}, for nearest-neighbors-only graph representation (NGR) module, which encodes the crystalline structures into crystal graphs, only the chemical bonds of closest neighboring atoms were considered. 
        Each node in the graph is an atom in the crystal lattice.
        Each edge in the graph is a chemical bond between nearest neighbors.
        Here, ``nearest neighbors" means a set of neighboring atoms for a centering atom.
        There is an edge between two neighboring atoms when $d_{\rm{neighbor}}$ $<$ 1.2 * $d_{\rm{nearest}}$, where $d_{\rm{neighbor}}$ is the distance between those two neighboring atoms, $d_{\rm{nearest}}$ is the distance to the nearest neighbor for either of those two atoms.
        There are 16 features (various atomic attributes) on each node and one feature (normalized bond length) on each edge (as shown Tab.~\ref{tab: node features}).
        

\begin{table}[htbp] \small
    \caption{ 
        Node features.
    \hspace*{\fill}}
    \label{tab: node features}
    \begin{tabular*}{\hsize}{
        @{\extracolsep{\fill}} 
        ll
        @{\extracolsep{\fill}}
    }        
    \toprule
        No. &
        feature name \\          
    \midrule
        1 &
        atomic mass $^{a}$ \\
        2 &
        vdW radius $^{a}$ \\
        3 &
        atomic radius $^{a}$ \\
        4 &
        electronegativity (Pauling) $^{b}$ \\
        5 &
        orbital energy of highest valence electrons $^{c}$ \\
        6 &
        orbital energy of lowest valence electrons $^{c}$ \\
        7 &
        number of unfilled valence orbitals \\
        8 &
        number of filled valence orbitals \\
        9 &
        number of unfilled valence $s$ orbitals \\
        10 &
        number of filled valence $s$ orbitals \\
        11 &
        number of unfilled valence $p$ orbitals \\
        12 &
        number of filled valence $p$ orbitals \\
        13 &
        number of unfilled valence $d$ orbitals \\
        14 &
        number of filled valence $d$ orbitals \\
        15 &
        number of unfilled valence $f$ orbitals \\
        16 &
        number of filled valence $f$ orbitals \\
    \bottomrule
    \end{tabular*}
    \\[6pt]
        $^{a}$ {https://github.com/materialsproject/pymatgen} 
        
        $^{b}$ {https://www.webelements.com/} 

        $^{c}$ F. Herman, \textit{et al.}, Atomic Structure Calculations. Vol. 111, Prentice Hall, 1964. {doi: 10.1149/1.2426131}
        
\end{table}


\section*{3. Results and discussion}

    \subsection*{3.1 Modeling}
    
\begin{figure}[!tb]
            \centering
            \includegraphics[width = 7.2 cm]{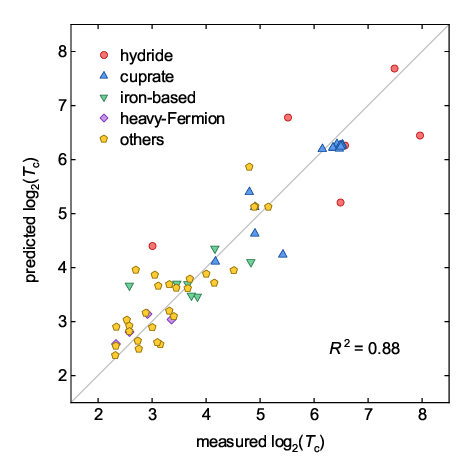}
            \caption{
            Model performance.
            Colored points: test data (randomly 10\% data).
            }
            \label{f2}
        \end{figure}

        Regression models predicting $T_{\rm{c}}$ (actually $log_{2}T_{\rm{c}}$) were trained and tested with the split datasets. 
        Figure~\ref{f2} shows the result of the best one in five models.
        Those five models achieved an average predictive score of $R^{2}$ = 0.85 $\pm$ 0.05. 
        

        
\begin{table}[htbp] \small 
    \caption{ 
        Three superconductors discovered recently (not in train data).
    \hspace*{\fill}}
    \label{tab: discovered recently}
    \begin{tabular*}{\hsize}{
        @{\extracolsep{\fill}} 
        llll
        @{\extracolsep{\fill}}
    }        
    \toprule
        No. & 
        material &
        $T_{\rm{c}}^{\rm{exp}}$ (K) &     
        $T_{\rm{c}}^{\rm{pred}}$ (K) \\          
    \midrule
        1 & CaH$_{6}$ @172 GPa & 215 & 249 \\
        2 & Ti @248 GPa & 26 & 16 \\
        3 & CsV$_{3}$Sb$_{5}$ & 2.3 & 11 \\
    \bottomrule
    \end{tabular*}
\end{table}   

        
        Table~\ref{tab: discovered recently} shows the prediction on three superconductors found in recent few years, which are not in the input data. 
        It can be seen that our GNN model has a good performance, not only in predicting various superconducting materials, but also in predicting unexplored materials (unseen data to the model).
        
    \subsection*{3.2 Screening high temperature superconductors} 

        The model shown in Fig.~\ref{f2} was used to predict the materials in ICSD.
        The ICSD contains over $200,000+$ entries of ordered or disordered crystal structures.
        Here we take all $110,000+$ ordered structures as the data to predict. 
        Table~\ref{tab: top 10 by model} shows the top 10 (higher $T_{\rm{c}}$) candidates of HTS, proposed by our GNN model.
        
        
\begin{table}[htbp] \small
    \caption{ 
        Top 10 HTS candidates (highest $T_{\rm{c}}^{\rm{pred}}$), proposed by our GNN model.
    \hspace*{\fill}}
    \label{tab: top 10 by model}
    \begin{tabular*}{\hsize}{
        @{\extracolsep{\fill}} llll @{\extracolsep{\fill}}
    } 
    \toprule
        No. & 
        material &
        $T_{\rm{c}}^{\rm{pred}}$ (K) &
        category \\ 
    \midrule
        1 & CF$_{4}$ & 271 & molecular crystal \\ 
        2 & CaZnF$_{4}$ & 270 & ionic crystal \\ 
        3 & C$_{2}$F$_{6}$O$_{3}$ & 270 & molecular crystal \\ 
        4 & NH$_{4}$F & 269 & ionic crystal \\ 
        5 & MgF$_{2}$ & 265 & ionic crystal \\ 
        6 & LiCaF$_{3}$ & 262 & ionic crystal \\ 
        7 & AlF$_{3}$ & 254 & ionic crystal \\ 
        8 & CaO$_{2}$ & 254 & ionic crystal \\ 
        9 & CaF$_{2}$ & 250 & ionic crystal \\ 
        10 & CaCO$_{3}$ & 248 & ionic crystal \\ 
    \bottomrule
    \end{tabular*}
\end{table}   
        
        
        \begin{figure}[!tb]
            \centering
            \includegraphics[width = 7.2 cm]{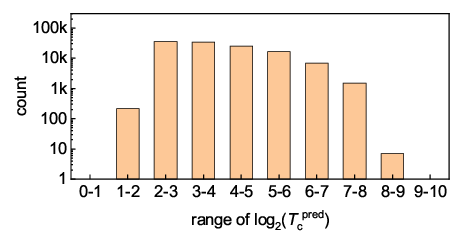}
            \caption{
            Statistical chart of the predictions on the materials in ICSD.
            }
            \label{f3}
        \end{figure}
        
        Unexpectedly, we found that nearly half of the predicted $T_{\rm{c}}$ is higher than 30 K, which is inconsistent with the domain knowledge. 
        Of course, it doesn't mean our predictions are unreliable.
        It just means that our model can not directly predict new high temperature superconductors, because the regression model can only predict how high the $T_{\rm{c}}$ could be, but can not predict whether or not the material is a superconductor. 
        More specifically, the model often mistakes insulators as high temperature superconductors. 
        But, many materials in ICSD are not superconductors, even not metals, but insulators.
        The predicted $T_{\rm{c}}$ of insulators should be seen as the $T_{\rm{c}}$ they could have, if they could be changed into metals by chemical doping or high pressure.
        Unfortunately, most insulators would not change into metals, no matter how they are doped or compressed. 
        They may have met the necessary conditions for high $T_{\rm{c}}$, but just don't have the chance to be superconductors.
        So, we need a materials screening to rule out the unchangeable insulators (mostly ionic crystals and molecular crystals). 
        
        Here, we used the data of energy gap (given by DFT calculation) in MATGEN \cite{MATGEN} to identify insulators.
        The materials $E_{g}$ $>$ 2 eV were removed from the candidates list.
        Then, we further screened the insulators in the HTS candidate list manually. 
        The ionic crystals and molecular crystals were removed from the candidates list.
        Table~\ref{tab: top 10 after manual screening} shows the top 10 candidates after manual screening. 
        
        
\begin{table}[htbp] \small
    \caption{ 
        Top 10 HTS candidates after manual screening (may need further doping).
    \hspace*{\fill}}
    \label{tab: top 10 after manual screening}
    \begin{tabular*}{\hsize}{
        @{\extracolsep{\fill}} llll @{\extracolsep{\fill}}
    } 
    \toprule
        No. & 
        material &
        $T_{\rm{c}}^{\rm{pred}}$ (K) &
        ICSD ccode \\ 
    \midrule
        1	&	CaB$_{6}$O$_{10}$                &	154  &	161320	\\
        2	&	Ca$_{2}$Ga$_{2}$O$_{5}$          &	145  &  051545	\\
        3	&	CaC$_{2}$Bi$_{2}$O$_{8}$         &	138  &  094741	\\
        4	&	Co$_{0.89}$Ca$_{1.11}$GeO$_{4}$  &	136  &  173465	\\
        5	&	Sn$_{0.976}$Au$_{0.024}$F$_{4}$	 &	132  &  078909	\\
        6	&	Ca$_{2}$B$_{5}$O$_{9}$Br         &	126  &  018001	\\
        7	&	Ca$_{2}$Nb$_{2}$O$_{6}$F         &	121  &  063189	\\
        8	&	RbNi$_{2}$F$_{6}$      	         &	113  &  031781	\\
        9	&	LiB$_{6}$O$_{9}$F                &	111  &	420286 	\\
        10	&	YCaGa$_{3}$O$_{7}$   	         &	103  &  109448	\\
    \bottomrule
    \end{tabular*}
\end{table}   

        
    \subsection*{3.3 Understanding the GNN model from the predictions} 
        
        In order to understand what the model have learn, we looked for the patterns in the predictions.
        For that purpose, there is no need to box ourselves into the materials experimentally observed. 
        So, we conceived a series of binary compounds, with the sphalerite structure (space group: F$\overline{4}$3m, No.216), and in different lattice parameters. 
        We choose the structure of sphalerite because in this structure, there is only one type of chemical bond (for example, Ga-As bond is the only one in GaAs, and the Ga-Ga and As-As is too far to form a substantial bond). 
        Then the $T_{\rm{c}}$ of these crystals of binary compound (CBC), if exists, could be considered to be solely contributed by that bond. 
        They are purest systems to see the influence of chemical bond on the superconductivity.
        
        Most of these CBC are not superconducting, or not metallic, or even thermodynamically unstable. 
        But by predicting these materials don't exist, we can see the influence of bond length and elements to the $T_{\rm{c}}$.    
        The meaning of the predictions of these CBC is how high their $T_{\rm{c}}$ maxima could be, if they were superconductors. 
        
\begin{figure*}[!tb]
            \centering
            \includegraphics[width = 16 cm]{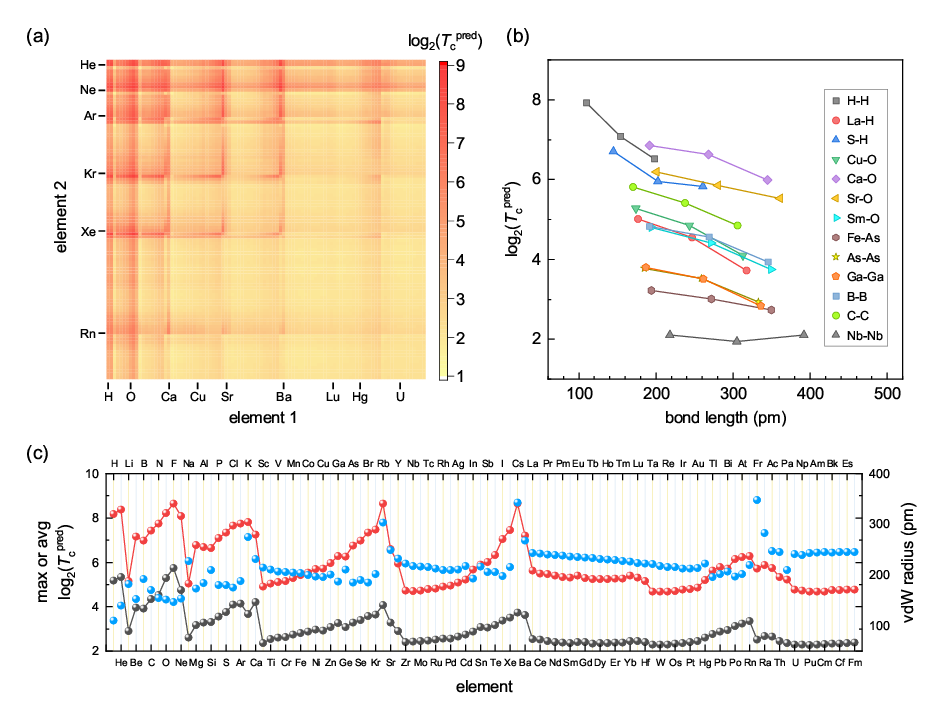}
            \caption{
            Results of predicting conceived binary compounds. 
            (a) Each $T_{c}^{pred}$ value represents the predicted $T_{c}$ of the conceived binary compounds consisting of two chemical elements (marked as ``element 1" and ``element 2"). 
            And the bond length between the two nearest neighboring atoms is 50\% of the sum of van der Waals (vdW) radii ($d = r_{1}^{vdW} + r_{2}^{vdW}$).
            (b) Dependence of $T_{c}$ on bond length. The three data points in each line are 50\%, 70\%, and 90\% of $d$, respectively.
            (c) Black and red: the average / maximal predicted $T_{c}$ for each element, derived from subplot (a), blue: the vdW radius of each element.
            }
            \label{f4}
        \end{figure*}

        Figure~\ref{f4}a shows a result of the predictions of CBC. Each $T_{\rm{c}}$ value in Fig.~\ref{f4}a represents the theoretical $T_{\rm{c}}$ maxima of the CBC consisting of each two chemical elements in the heat map. 
        It can be found that the $T_{\rm{c}}$ varies with the combination of elements. 
        The chemical elements of the two atoms forming the bond (not each single element, but the element combination) is vital for high $T_{\rm{c}}$. 
        Some element combinations are able to support high $T_{\rm{c}}$, whereas some others never had a chance.

        Meanwhile, Fig.~\ref{f4}b shows that shorter bonds usually lead to higher $T_{\rm{c}}$, which is consistent with the common sense, and indicating that our model do have learnt the dependence of $T_{\rm{c}}$ on bond length. 
        On the basis of that, we can predict the $T_{\rm{c}}$ of superconductors at high pressure, with confidence.
        
        Please note that what is shown in the heat-map (Fig.~\ref{f4}a) is not the contribution of each element to the $T_{\rm{c}}$, but of each combination of two elements. 
        Here, it's inappropriate to discuss either one element alone.
        However, still, we can see the differentiation between elements by calculate the maximal or average value of each row or column in the heat-map, as shown in Fig.~\ref{f4}c. 
        It suggests that some elements are more favorable than others for higher $T_{\rm{c}}$. 
        Especially, Alkali, Alkaline earth, Chalcogen, and Halogen elements are preferred.
        
        According to the domain knowledge of superconductivity, the factors affecting $T_{\rm{c}}$ are numerous. 
        Bond length and elements combination are only two of them. 
        Having learnt the influence of bond length, our model has surpassed all previous ML models, but still not goes beyond human expertise.
        Meanwhile, the other knowledge that our model has learnt, the influence of elements combination, is a new information that has never been noticed and interpreted.
        
        
    \subsection*{3.4 Hunting for superior superconductors in compressed hydrides}
        
\begin{figure}[!tb]
            \centering
            \includegraphics[width = 8 cm]{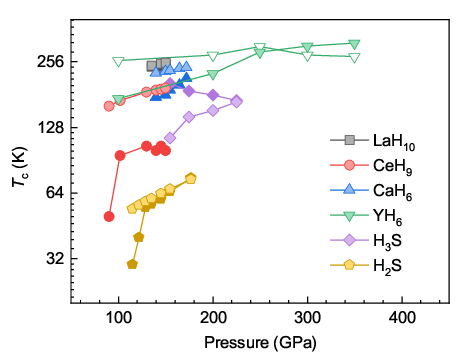}
            \caption{
            Variation of predicted $T_{c}$ with pressure for several typical compressed hydrides. 
            Solid filled points: experimental values, 
            light filled points: predicted values, 
            open filled points: DFT calculated values. 
            }
            \label{f5}
        \end{figure}

        In addition, since our model has learnt the variation of $T_{\rm{c}}$ with bond length, it is supposed to get good performance in predicting superconductors at high pressure. 
        So we used the model shown in Fig.~\ref{f2} to predict compressed hydrides.
        Figure~\ref{f5} shows the results of several typical hydrides, in which YH$_{6}$, LaH$_{10}$, H$_{3}$S and H$_{2}$S are in train data, whereas CeH$_{9}$ and CaH$_{6}$ are unseen data. 
        On this basis, our model can be used to search new compressed hydrides with higher $T_{\rm{c}}$ or lower pressure. 

        
\section*{4. Conclusion} 

    In this study, a bond sensitive GNN model (BSGNN) was developed to predict the $T_{\rm{c}}^{\rm{max}}$ of various superconducting materials. 
    Our work suggests that a well-designed GNN algorithm can achieve rational, reliable and useful results on complex materials problems with unclear mechanism, such as superconductivity. 
    Owing to the GAT and CMP techniques, BSGNN has effectively learnt the dependence of $T_{\rm{c}}$ on both bond length and chemical composition.
    The predictions of BSGNN shows that shorter bond length, as well as specific elements with relatively larger vdW radii, is in favor of high $T_{\rm{c}}$. 
    It illustrates the necessity and importance of considering crystalline structure information when predicting superconductors. 
    As the influence of bond length was learnt, BSGNN shall have advantages in predicting superconducting materials under different pressure. 
    With the help of BSGNN, we predicted all the materials in ICSD, and proposed some promising candidates of HTS after further artificial screening.

\section*{Acknowledgements}

    The authors gratefully acknowledge the financial support of National Key Research and Development Program of China (2022ZD0117805) and Guangdong Province Key Area Research and Development Program (2019B010940001). 

\section*{Supplementary material}

    Supplementary material for this article is available \href{https://github.com/GLinustb/BSGNN}{online}.
    
\section*{Conflict of interest}

    The authors declare no competing interests.
    
\section*{ORCID iDs}

    \noindent Yang Liu \href{https://orcid.org/0000-0001-6683-7610}{https://orcid.org/0000-0001-6683-7610}
    
    \noindent Pin Chen \href{https://orcid.org/0000-0001-8746-9917}{https://orcid.org/0000-0001-8746-9917}
    
    \noindent Haiyou Huang \href{https://orcid.org/0000-0002-2801-2535}{https://orcid.org/0000-0002-2801-2535}
    
    \noindent Yanjing Su \href{https://orcid.org/0000-0003-2773-4015}{https://orcid.org/0000-0003-2773-4015}


\bibliography{references} 

\end{document}